\documentclass[twocolumn,showpacs,amsmath,amssymb]{revtex4-1}
\usepackage{graphicx}
\usepackage{epstopdf}

\begin{document}

\title{Conductance enhancement due to atomic potential fluctuations in graphene}
\author{Dima Bolmatov$^{1}$\footnote{e-mail: d.bolmatov@qmul.ac.uk}}
\author{D. V. Zavialov$^{2,3}$}
\address{$^1$ School of Physics, Queen Mary University of London, Mile End Road, London, E1 4NS, UK}
\address{$^2$ Volgograd State Social-Pedagogical University, Volgograd, 400005, Russia}
\address{$^3$ Volgograd State Technical University, Volgograd, 400005, Russia}
\begin{abstract}
We solve the Dirac equation, which  describes charge massless chiral relativistic carriers  in a two-dimensional graphene. We have identified and analysed a novel pseudospin-dependent scattering effect. We compute the tunneling conductance and generalize the analytical result in the presence of the tunable atomic potential of a graphene strip. The absence of back scattering in graphene is shown to be due to Berry's phase which corresponds to a sign change of the wave function under a spin rotation of a particle. We use the transfer matrix approach and find that the electric conductance of doped graphene increases due to atomic potential fluctuations.

\end{abstract}
\maketitle
\section{Introduction}
Graphene is a single layer of carbon atoms densely packed in a honeycomb lattice. This material was found in its free state only recently, when individual graphene samples of a few microns in size were isolated by micromechanical cleavage of graphite \cite{Novoselov-1}. The current intense interest in graphene is driven by both the unusual physics involved and a realistic promise of device applications. Among relativistic like phenomena observed in graphene so far, there are two new types of the integer quantum Hall effect and the presence of minimal metallic conductivity of about one conductivity quantum\cite{Slon-1}. Transport properties of graphene are interesting because of their unique topological structure \cite{Voz-1,Bolmat-1,bolmat-2,Roy-1}. There have been some reports on experimental study of transport in graphene \cite{Geim-1}. Various calculations have been performed to understand energy bands of graphene \cite{Shyt-1,Loz-1}. It has been successful in the study of various properties including the Aharonov-Bohm  effect \cite{Morp-1}, optical absorption spectra \cite{MKatsnelson-1}, quantum tunneling in graphene-based structures \cite{bolmat-3,zavialov-1,bolmat-4}, ferromagnetism \cite{ming-1} and instabilities in the presence and absence of a magnetic field \cite{Loz-2}.

In a two-dimensional world, there are two basic motions: forward and backward. Random scattering can cause them to mix, which leads to resistance. Just as we have learned from the basic traffic control, it would be much better if we could spatially filter the counterflow directions. This effect could be used to produce a valley-polarized current out of an unpolarized stream of electrons due to line defect particularly in graphene \cite{Gunlycke-1, feng-1, Gun-2, golub-1,ram-1,liw-1,feng-2,chi-1,Leo-1,hy-1}.

The electronic transport properties of graphene with defects and impurities exhibit pronounced differences from those of conventional two dimensional electron systems investigated in the past. The basic question that we seek to answer is what happens to the tunneling  as we approach the Dirac point of zero carrier concentration with the presence of electrically charged scatters.
\begin{figure}
	\centering
\includegraphics[scale=0.24]{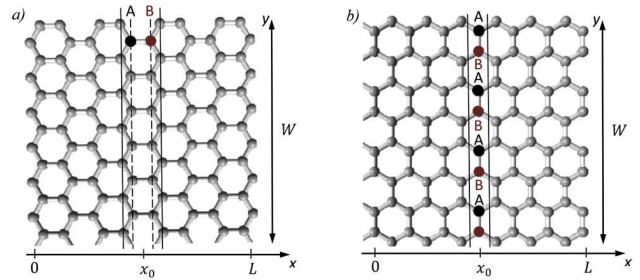} 
\caption{Schematic of a graphene layer with a line tunable potential. The line potential is placed inside the graphene strip enhances the conductance in a vicinity of the Dirac point $\epsilon\ll\hslash v/L$, where $L$ is the length of the strip with the width $W$. The indices $A$, $B$ label the two sublattices of the honeycomb lattice of carbon atoms has different atomic potentials, which can be tuned.}
	\label{fig-1}
\end{figure}
In the present work we consider the effects of atomic potential fluctions in the setup depicted in fig. 1. At low doping the conductance is determined by quasiparticle tunneling, which is
independent of the boundary conditions in the $y$-direction if $L\ll W$. 

\section{back scattering in graphene}
\subsection{General consideration}
The quasiparticle excitation spectrum of the graphene-based junction consists of the positive eigenvalues of the Dirac  equation. The Dirac equation has the form of two equations for electron $u(\bold{r})_{A,B}$ and hole wavefunctions $v(\bold{r})_{A,B}$. In this study we restrict ourselves to the single-valley
Dirac equation for graphene
\begin{eqnarray}\label{Diraceq}
-i\hbar v_{F}\bold{\sigma}\bold{\nabla}u+ Vu=\varepsilon u
\end{eqnarray}
where $u$ is a spinor of wave amplitudes for two nonequivalent sites of the honeycomb lattice. The Fermi energy $\varepsilon$ and the impurity potential V (x, y) in the
graphene sample ($0<x<L$) are considered to be much smaller than the Fermi energy $E_{F}$ in the ideal metallic leads ($x<0$ and $x>L$). For zero doping the conductance
is determined by the states at the Dirac point, $\varepsilon=0$. Transport properties at finite energies determine the conductance of doped graphene.

Since for aspect ratios $W\gg L$ the boundary conditions in the $y$-direction are irrelevant, so we take periodic boundary conditions for simplicity. Different wave vectors  $q_{n}$=2$\pi$n/W (with n=0, $\pm$ 1, $\pm$ 2, \ldots) in the $y$-direction  parallel to the interfaces(at the points $x=0$ and $x=L$) and not coupled, so we can consider each transverse mode separately. Without impurity in the bulk at a given energy $\varepsilon$ and transverse wave vector $q$ we have up to two basis states
\begin{eqnarray}\label{electron}
u^{\pm}(\bold{r})=\frac{1}{\sqrt{cos{\alpha}}}\chi_{n}(y)\exp^{\pm ikx}\left( \begin{array}{c}
\begin{array}{cc}
\exp^{\mp i\alpha/2} \\ \pm\exp^{\pm i\alpha/2}
 \end{array}
 \end{array}
 \right)
\end{eqnarray}
with the definitions
\begin{eqnarray}\label{angle}
&&	\alpha(\epsilon)=\arcsin[\hbar v_{F} q/(\epsilon+E_{F})], \\ 	
&& k(\epsilon)=(\hbar v_{F})^{-1}(\epsilon+E_{F})\cos(\alpha)	
\end{eqnarray}
 The angle $\alpha$ $\in$ (-$\pi$/2,$\pi$/2) is the angle between  the
initial and final wave vectors $\bold{k}_{i}=(k,q)\vert_{x=0}$ and $\bold{k}_{f}=(k,q)\vert_{x=L}$ . With this sign convention the state $u^{+}$  move in the $+x$, while $u^{-}$ move in the $-x$ direction. The factor $1/\sqrt{cos{\alpha}}$  ensure that two states carry the same particle current.

Considering a back scattering process $\bold{k}\rightarrow - \bold{k}$ in an arbitrary external potential we will confine ourselves to states in the vicinity of the $K$ point, but the extension to states near a $K^{'}$ point is straightforward. We have $\alpha(\bold{k})=0$ and  $\alpha(\bold{-k})=\pi$. Rotation in spin space can be obtained through the replacement $(s_{i},\bold{k}_{i})\rightarrow(s_{f},\bold{-k}_{f})$ corresponding to the electron motion of a time-reversal path. When the wave vector $\bold{k}$ is rotated in the anticlockwise direction adiabatically as a function of time $t$ around the origin for a time interval $0<t<T$, the wavefunction is changed into $u_{s}(\bold{k})\exp(-i\phi)$, where  $u_{s}(\bold{k})$ is the "spin" part of an eigenfunction of the Eq. \ref{Diraceq} and $\phi$ is Berry's phase. Choosing $\phi(\bold{k})=\alpha(\bold{k})/2$ in such a way that the wave function becomes continuous as a function of $\alpha(\bold{k})/2$ and given by
\begin{eqnarray}
\phi=-i\int_{0}^{T} dt\langle u_{s}(\bold{k}(t))\mid \frac{d}{dt} u_{s}(\bold{k}(t)) \rangle=\pi
\end{eqnarray}
This shows that the rotation in the $\bold{k}$ space by $2\pi$ leads to the change in the phase by $+\pi$, i.e., a sign change. A similar phase change of back scattering process is the origin of the so-called anti-localization effect in systems with strong spin-orbit scattering\cite{Larkin-1}. This anti-localization effect was observed experimentally \cite{Kom-1,Ber-1}. The absence of back scattering can be destroyed by various effects. When the potential range becomes shorter than the lattice constant, the back scattering appears two reasons. The first is the appearance of inter-valley matrix elements between $K$ and $K'$ points \cite{And-1}. The second, which we investigate in this work, is that the effective potential of an impurity for $A$ and $B$ sites in honeycomb lattice can be different. This causes mixing of the spin space and the momentum space. 
\subsection{Long range impurities}
There is an absence of back scattering for long range impurities. This can be
seen easily: using the formalism of the previous subsection we can simply show that
the matrix element of the back-scattering process is zero
\begin{eqnarray}\label{imp}
\nonumber \langle -\bold{k}\vert H_{imp}\vert\bold{k}\rangle=\frac{1}{2S}\int d\bold{r}e^{i2\bold{k}\cdot\bold{r}}\left(e^{i\alpha_{-\bold{k}}/2}e^{-i\alpha_{-\bold{k}}/2} \right)\cdot \\
\left( \begin{array}{c}
\begin{array}{cc}
V(\bold{r}) & 0 \\ 0 & V(\bold{r})
 \end{array}
 \end{array}
 \right)
 \left( \begin{array}{c}
\begin{array}{cc}
e^{-i\alpha_{\bold{k}}/2} \\  e^{i\alpha_{\bold{k}}/2}
 \end{array}
 \end{array}
 \right)=0
\end{eqnarray}

(here $S$ is the unit cell area of the graphene lattice). This is rather dramatic
since all real materials have impurities, and it is normally these which determine
the transport properties of the material. However, in the case of graphene the
impurities don't scatter, i.e. they are effectively not there, and hence one has
the possibility of coherent transport in graphene.

\begin{figure*}
\begin{center} 
	\centering
\includegraphics[scale=0.57]{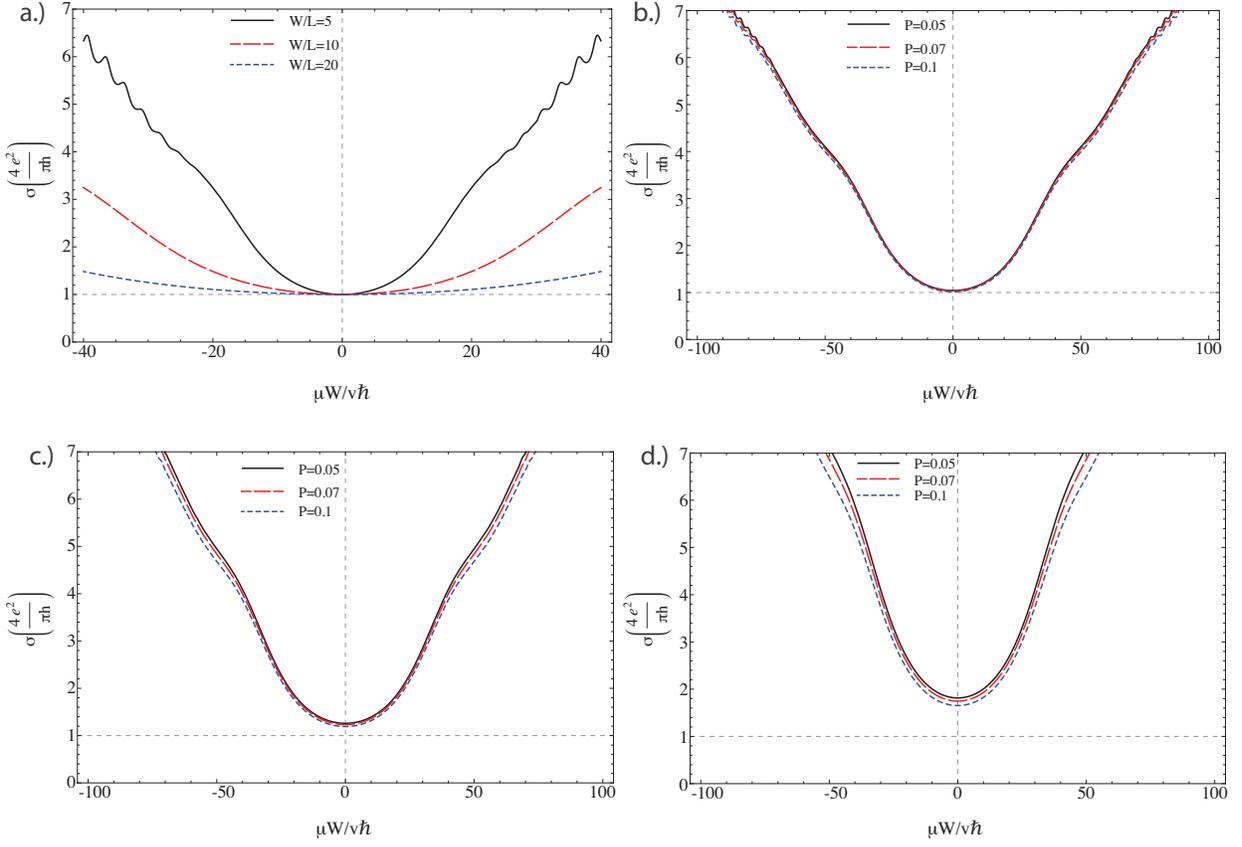} 
\end{center}
\caption{Gate voltage dependence of the conductance (calculated from Eq.\ref{con}) for the case of a smooth edge $q_{n}=\left( n+\frac{1}{2}\right)\frac{\pi}{W}$: a.) for a different aspect ratios $W/L=5$, $W/L=10$, $W/L=20$, and for incident angle $\alpha=0^{\circ}$; b.), c.), d.) for a fixed aspect ratio $W/L=10$ and for $\alpha=30^{\circ}$, $\alpha=60^{\circ}$, $\alpha=90^{\circ}$ angles correspondingly. The horizontal asymptotes are indicated by dashed lines.}
	\label{fig-1}
\end{figure*}

\section{transfer matrix approach}
We first determine the transfer matrix $M_{n}(x,0)$ of the $n$th mode $u_{n}(x)e^{iq_{n}y}$ through the undoped graphene ribbon. The evolution of $u_{n}(x)$ inside the graphene sample can be written as $u_{n}(x)=M_{n}(x,0)u_{n}(0)$, where the transfer matrix $M_{n}(x)$ should satisfy a generalized unitarity condition
\begin{eqnarray}\label{matrix}
 M^{-1}=\sigma_{x}M^{\dagger}\sigma_{x} 
\end{eqnarray}
In general Eq.\ref{imp} (ignoring an irrelevant scalar phase factor) restricts $M$ to a three-parameter form
\begin{eqnarray}
 M=e^{i\alpha_{1}\sigma_{z}}e^{i\alpha_{2}\sigma_{y}}e^{i\alpha\sigma_{x}}
\end{eqnarray}
The real parameters $\alpha_{1}$, $\alpha_{2}$, $\alpha$ depend on the boundary at the scale of the lattice constant and they cannot be determined from the
 Dirac equations. In this work we consider the ideal interfaces $x=0$ and $x=L$ case, when $\alpha_{1}=0=\alpha_{2}$. From the Dirac  Eq.\ref{Diraceq} we obtain the differential equation
\begin{eqnarray}
\frac{d}{dx}M_{n}(x,0)=\left( \frac{i\varepsilon}{\hbar v}\sigma_{x}+q_{n}\sigma_{z}\right) ,
\end{eqnarray} 
with solution for $V(x,y)=0$
\begin{eqnarray}
M_{n}(x,0)=\cos{k_{n}x}+\frac{\sin{k_{n}x}}{k_{n}}\left( \frac{i\varepsilon}{\hbar v}\sigma_{x}+q_{n}\sigma_{z}\right)
\end{eqnarray}
where $k_{n}=\sqrt{\left((\frac{\varepsilon}{\hbar v}\right)^{2}-q_{n}^{2}}$ is the longitudinal wave vector, $q_{n}$ is the transversal wave vector. The total transfer matrix through the impurity-free graphene ribbon ($0<x<L$) is
\begin{eqnarray}
M_{tot}=M(L,0)e^{i\alpha\sigma_{x}}M(0,0)
\end{eqnarray}
Now we construct the transfer matrix for the electron (hole) quasiparticle excitations in the graphene ribbon for disorder potential case, where $V$ is the charged impurity potential given by a diagonal matrix, i.e.,
\begin{eqnarray}\label{pot}
V=
\left( \begin{array}{c}
\begin{array}{cc}
\gamma_{A}\delta(x-x_{0}) & 0 \\ 0 & \gamma_{B}\delta(x-x_{0})
 \end{array}
 \end{array}
 \right)
\end{eqnarray}
where $\gamma_{A}$ ($\gamma_{B}$) is the corresponding microscopic potential and
smooth on atomic scales, which is localized along a line $x=x_{0}$ and placed in the zone corner $K$ ($K^{'}$) of sublattice $A$ ($B$). The Fermi-energy $\varepsilon$ and impurity potential $V$ (Eq.\ref{pot}) in graphene sample ($0<x<L$) are considered to be
much smaller than the Fermi-energy $E_{F}$ in the ideal metallic leads ($x<0$ and $x>L$). For zero doping the conductance is determined by the states at the Dirac point, $\varepsilon=0$. Transport properties at finite energies determine the conductance of doped graphene.

The great physical and analytical advantage of the $M$ matrix is that it is $multiplicative$. In order to obtain transfer matrix for disorder region (along line $x=x_{0}$) graphene $M_{d}$ we integrate Eq. \ref{Diraceq}
\begin{eqnarray}
\nonumber
\lim_{\nu\rightarrow 0}\int^{x_{0}+\nu}_{x_{0}-\nu}\left(-i\hbar v_{F}\sigma\cdot\bigtriangledown u +V u  \right)dx=\lim_{\nu\rightarrow 0}\int^{x_{0}+\nu}_{x_{0}-\nu} \varepsilon u dx
\end{eqnarray}
Making use of well-known algebraic property for Pauli matrices
\begin{eqnarray}
\sigma^{2}_{x}=\sigma^{2}_{y}=\sigma^{2}_{z}=-i\sigma_{x}\sigma_{y}\sigma_{z}=
\left( \begin{array}{c}
\begin{array}{cc}
1 & 0 \\ 0 & 1
 \end{array}
 \end{array}
 \right)
\end{eqnarray}
we easily obtain the final form for total transfer matrix $M$ in disorder graphene sample
\begin{eqnarray}
 M_{tot}=M(L,0)e^{i\frac{\alpha}{2}\sigma_{x}}M_{d}e^{i\frac{\alpha}{2}\sigma_{x}}M(0,0)
\end{eqnarray} 
where
\begin{eqnarray}\label{matrix}
M_{d}=\left( \begin{array}{c}
\begin{array}{cc}
 1 &  \tilde{\gamma}_{A} \\ \tilde{\gamma}_{B} & 1
 \end{array}
 \end{array} \right)
\end{eqnarray}
and $\tilde{\gamma}_{A(B)}=-i\gamma_{A(B)}/\hbar v_{F}$.
The transfer matrix of the whole sample is straightforwardly related to the matrices of transmission and reflection amplitudes
\begin{eqnarray}
M\equiv M_{tot}=
\left( \begin{array}{c}
\begin{array}{cc}
1/t^{\dagger} & -r^{\dagger}/t^{\dagger} \\ -\tilde{r}/\tilde{t} & 1/\tilde{t}
 \end{array}
 \end{array}
 \right)
\end{eqnarray}
The conductance of the graphene strip is expressed through the transmission amplitudes
by the Landauer formula 
\begin{equation}\label{con}
G=g_{0}\sum_{n=0}^{N-1}Tn, \ \ g_{0}=4e^{2}/h
\end{equation}
where $T=Tr[t^{\dagger}t]$. At the Dirac point $V_{gate}=0$ the transmission probability for the case $N\gg W/L$ reads 
\begin{equation}\label{prob}
T_{n}=\frac{1}{\left(\cos{\frac{\alpha}{2}}+P\sin{\frac{\alpha}{2}}\right)^{2}}\frac{1}{\cosh^{2}{\left[\pi(n+\frac{1}{2})\frac{L}{W}\right]}}
\end{equation}
When parameters $P=0$ and $\alpha=0$ the Eq.\ref{prob} reduces to equation $(4)$, derived in the paper \cite{Ben-1}. Parameter $P=\mid\gamma_{A}-\gamma_{B}\mid$ reperesents  atomic potential fluctuations, when finite $P\ll \hbar\upsilon_{F}$.
\section{Conclusion}
In this paper we used the transfer matrix approach to calculate the  tunneling conductance. We generalize the analytical result in the presence of the tunable atomic potential of a graphene strip and find that the electric
conductance of doped graphene increases due to atomic potential fluctuations. The absence of back scattering in graphene is shown to be due to Berry's phase which corresponds to a sign change of the wave function under
a spin rotation of a particle. We have identified and analysed a novel pseudospin-dependent scattering
effect. The low cost experimental techniques allow for easy verification of the proposed hypothesis.
\section{Acknowledgements}
We are indebted to Ben Still for assistance. D. B. thanks Myerscough Bequest for financial support. D. B. acknowledges Thomas Young Centre for Junior Research Fellowship and Cornell University (Neil Ashcroft and Roald Hoffmann) for hospitality.


\begin{thebibliography}{99}

\bibitem{Novoselov-1} K. S. Novoselov, A. K. Geim, S. V. Morozov, D. Jiang, M. I. Katsnelson, I. V. Grigorieva, S. V. Dubonos, and A. A. Firsov, Nature {\bf 438}, 197 (2005).

\bibitem{Slon-1} J. C. Slonczewski and P. R. Weiss, Phys. Rev. {\bf 109}, 272 (1958); G. W. Semenoff, Phys. Rev. Lett. {\bf 53}, 2449 (1984); F. D. M. Haldane, Phys. Rev. Lett. {\bf 61}, 2015 (1988); J. Gonzalez, F. Guinea, and M. A. H. Vozmediano, Nucl. Phys. {\bf B406}, 771 (1993);

\bibitem{Voz-1} Fernando de Juan, A. Cortijo, and Mara A. H. Vozmediano, Phys. Rev. B {\bf 76}, 165409 (2007).


\bibitem{Bolmat-1} D. Bolmatov and C.-Y. Mou, Physica B: Condensed Matter {\bf 405}, 2896 (2010).

\bibitem{bolmat-2}  D. Bolmatov, C.-Y. Mou, JETP {\bf 110}, 612 (2010).

\bibitem{Roy-1} J. K. Pachos, M. Stone, and K. Temme, Phys. Rev. Lett. {\bf 100}, 156806 (2008).

\bibitem{Geim-1} C. Casiraghi, A. Hartschuh, E. Lidorikis, H. Qian, H. Harutyunyan, T. Gokus, K. S. Novoselov, and A. C. Ferrari, Nano Lett. {\bf 7} (9), 2711 (2007).

\bibitem{Shyt-1} A. Shytov, M. Rudnerb, Nan Guc, M. Katsnelson and L. Levitov, Solid State Commun. {\bf 149}, 1087 (2009).

\bibitem{Loz-1} Yu. E. Lozovik and A. A. Sokolik, Phys. Lett. A {\bf 374}, 326 (2009).

\bibitem{Morp-1} S. Russo, Jeroen B. Oostinga, D. Wehenkel, H. B. Heersche, S. S. Sobhani, L. M. K. Vandersypen, and A. F. Morpurgo, Phys. Rev. B {\bf 77}, 085413 (2008).

\bibitem{MKatsnelson-1} M. I. Katsnelson, EPL {\bf 84}, 37001 (2008).

\bibitem{bolmat-3} D. Bolmatov and C.-Y. Mou, JETP {\bf 112}, 102 (2011).

\bibitem{zavialov-1} D. V. Zavialov, V. I. Konchenkov and S. V. Kruchkov, Semiconductors {\bf 46}, 109 (2012).

\bibitem{bolmat-4} D. Bolmatov, Physica C: Superconductivity {\bf 471}, 1651 (2011).

\bibitem{ming-1} B. Soodchomshoma, I-Ming Tanga, and R. Hoonsawat, Phys. Lett. A {\bf 372}, 5054 (2008).

\bibitem{Loz-2} O. L. Bermana, R. Ya. Kezerashvilia and Y. E. Lozovik, Phys. Lett. A {\bf 372}, 6536 (2008).

\bibitem{Gunlycke-1} D. Gunlycke and C. T. White, Phys. Rev. Lett. {\bf 106}, 136806 (2011).

\bibitem{feng-1} Feng Zhai and Kai Chang, Phys. Rev. B {\bf 85}, 155415 (2012).

\bibitem{Gun-2} Daniel Gunlycke, Carter T. White, Journal of Vacuum Science and Technology B Microelectronics and Nanometer Structures {\bf 30}, 03D112 (2012).

\bibitem{golub-1} L. E. Golub, S. A.  Tarasenko, M. V.  Entin, and L. I.  Magarill, Phys. Rev. B {\bf 84}, 195408 (2011).

\bibitem{ram-1} M. Ramezani Masir, A. Matulis, and F. M. Peeters, Phys. Rev. B {\bf 84}, 245413 (2011).

\bibitem{liw-1} Liwei Jiang, Xiaoling Lv, Yisong Zheng, Phys. Lett. A {\bf 376}, 136 (2011).

\bibitem{feng-2} Feng Zhai, Yanling Ma, Ying-Tao Zhang, J. Phys.: Condens. Matter {\bf 23}, 385302
(2011).

\bibitem{chi-1} Chi-Hsuan Chiu and Chon-Saar Chu, Phys. Rev. B {\bf 85}, 155444 (2012).

\bibitem{Leo-1} L. Mayrhofer and D. Bercioux, Phys. Rev. B {\bf 84}, 115126 (2011).

\bibitem{hy-1}  A. R. Wright and T. Hyart, Appl. Phys. Lett. {\bf 98}, 251902 (2011).

\bibitem{Larkin-1} S. Hikami, A.I. Larkin and Y. Nagaoka, Prog. Theor. Phys. {\bf 63}, 707 (1980).

\bibitem{Kom-1} F. Komori, S. Koboyashi and W. Sasaki, J. Phys. Soc. Jpn. {\bf 51}, 3162 (1982).

\bibitem{Ber-1} G. Bergmann, Phys. Rev. Lett. {\bf 48}, 1046 (1982).

\bibitem{And-1} T. Ando and T. Nakanishi, J. Phys. Soc. Jpn. {\bf 67}, 1704 (1998).

\bibitem{Ben-1} J. Tworzydlo, B. Trauzettel, M. Titov, A. Rycerz, and C. W. J. Beenakker, Phys. Rev. Lett. {\bf 96}, 246802 (2006).

\end{thebibliography}
\end{document}